\documentclass[aps,prl,twocolumn,showpacs,superscriptaddress,floatfix]{revtex4-1}
\usepackage{amsmath,amssymb}
\usepackage{graphicx}
\usepackage{epsfig}
\usepackage{color}
\usepackage{rotating}
\usepackage{bm}
\usepackage{setspace}
\usepackage{hyperref}

\begin{document}
\title{Experimental observation of pseudogap in a modulation-doped Mott insulator: Sn/Si(111)-($\sqrt{3}\times \sqrt{3}$)\textit{R}30$^\textrm{o}$}
\author{Yan-Ling Xiong}
\author{Jia-Qi Guan}
\author{Rui-Feng Wang}
\affiliation{State Key Laboratory of Low-Dimensional Quantum Physics, Department of Physics, Tsinghua University, Beijing 100084, China}
\author{\\Can-Li Song}
\email{clsong07@mail.tsinghua.edu.cn}
\affiliation{State Key Laboratory of Low-Dimensional Quantum Physics, Department of Physics, Tsinghua University, Beijing 100084, China}
\affiliation{Frontier Science Center for Quantum Information, Beijing 100084, China}
\author{Xu-Cun Ma}
\email{xucunma@mail.tsinghua.edu.cn}
\affiliation{State Key Laboratory of Low-Dimensional Quantum Physics, Department of Physics, Tsinghua University, Beijing 100084, China}
\affiliation{Frontier Science Center for Quantum Information, Beijing 100084, China}
\author{Qi-Kun Xue}
\affiliation{State Key Laboratory of Low-Dimensional Quantum Physics, Department of Physics, Tsinghua University, Beijing 100084, China}
\affiliation{Frontier Science Center for Quantum Information, Beijing 100084, China}
\affiliation{Beijing Academy of Quantum Information Sciences, Beijing 100193, China}
\affiliation{Southern University of Science and Technology, Shenzhen 518055, China}

\begin{abstract}
Unusual quantum phenomena usually emerge upon doping Mott insulators. Using a combinatorial molecular beam epitaxy system integrated with cryogenic scanning tunneling microscopy, we investigate the electronic structure of a modulation-doped Mott insulator Sn/Si(111)-($\sqrt{3}\times \sqrt{3}$)\textit{R}30$^\textrm{o}$. In underdoped regions, we observe a universal pseudogap opening around the Fermi level, which changes little with the applied magnetic field and the occurrence of Sn vacancies. The pseudogap gets smeared out at elevated temperatures and alters in size with the spatial confinement of the Mott insulating phase. Our findings, along with the previously observed superconductivity at a higher doping level, are highly reminiscent of the electronic phase diagram in the doped copper oxide compounds.
\end{abstract}

\maketitle
\begin{spacing}{1.000}
Mott insulators are materials in which constituent electrons cannot move freely because of strong electron repulsions and lie at the heart of strongly correlated electron physics. Upon chemical doping, the electrons delocalize and the Mott insulators can become conductive and even superconductive via a process well-known as Mott transition \cite{mott1961transition,mott1982review,belitz1994anderson}. Following this transition, many unusual phenomena emerge and have been largely explored in one archetypal example of the perovskite-type copper oxide compounds. These include, for example, the unconventional superconductivity with a record critical temperature ($T_\textrm{c}$) at ambient atmosphere \cite{lee2006doping}, the pseudogap (PG) phase above $T_\textrm{c}$ \cite{timusk1999pseudogap} and various different symmetry-broken states \cite{fischer2007scanning,damascelli2003angle,timusk1999pseudogap}. Among these phenomena, the PG phase stands out and continues to attract attention, because clarifying its microscopic origin from either the performed Cooper pairs \cite{renner1998pseudogap,chen2005bcs,geshkenbein1997superconductivity} or some competing orders \cite{loret2019intimate,Dynamical2017Caprara,kampf1990spectral} has been considered as an essential prerequisite to understanding the high-$T_\textrm{c}$ superconductivity in cuprates. After more than three decades of study, however, the PG phenomenology, which turns out to be more prominent in underdoped regions, remains one of the greatest mysteries in the high-$T_\textrm{c}$ cuprate superconductors.

Adsorption of metal atoms (such as Pb, In, and Tl) on  semiconductor substrates (e.g.\ Si(111)) often induces rich surface superstructures and intriguing electronic states at the two-dimensional (2D) limit, including superconductivity \cite{zhang2010superconductivity,Macroscopic2011Uchihashi,Unconventional2018Nakamura,ozer2006hard}, charge density waves \cite{matetskiy2019observation,carpinelli1996direct} and unusual magnetism \cite{zhou2017two}. Specifically, adsorption of one-third monolayer (ML) Sn on Ge(111) and Si(111) leads to the formation of a ($\sqrt{3}\times \sqrt{3}$)\textit{R}30$^\textrm{o}$ reconstruction known as $\alpha$ phase \cite{carpinelli1997surface,morikawa2002stm,ottaviano2000stm}. Such a triangular lattice, which is characteristic of half-filled dangling bond orbitals, forms a conceptually simple and alternative system to explore correlation related Mott physics in the 2D situation \cite{profeta2007triangular,modesti2007insulating,odobescu2017electronic}. Intriguingly, a superconducting state of $T_\textrm{c}$ $\sim$ 4.7 K has been recently observed in Sn/Si(111)-($\sqrt{3}\times \sqrt{3}$)\textit{R}30$^\textrm{o}$ (hereinafter referred to as $\sqrt{3}$-Sn) on chemical doping by employing heavily boron-doped Si substrates \cite{wu2020superconductivity}. Such a charge transfer mechanism has been well known as modulation doping in semiconductor heterostructures \cite{hsu1997electron,stormer1980two,people1984modulation} and recently evidenced in cuprates \cite{zhong2016nodeless,zhong2020direct}, thereby calling for further investigation of possible commonalities and distinctions between the doped $\sqrt{3}$-Sn and cuprates.

To date, however, only limited experiments were performed, especially leaving the electronic structure of $\sqrt{3}$-Sn in underdoped regions poorly understood. It remains a puzzle whether the opened energy gap at the Fermi level ($E_\textrm{F}$) from the tunneling experiment \cite{ming2018zero} relates to superconductivity, and if no why the Coulomb gap increases in magnitude with increasing doping concentration. Here we resolve this issue by combining a molecular beam epitaxy system (MBE) with cryogenic scanning tunneling microscopy (STM). We reveal the PG nature for the opened gap in the underdoped regions, which instead decreases with increasing doping. These behaviors bear a close resemblance to the established electronic phase diagram in cuprate superconductors.

All experiments were carried out in a commercial ultrahigh vacuum (UHV) STM apparatus (Unisoku USM 1300), integrated with an MBE chamber for \textit{in-situ} sample preparation. A magnetic field up to 11 T can be applied perpendicular to the sample surface. The base pressure for both chambers is better than 2.5 $\times$ 10$^{-10}$ Torr. Two distinct $p$-type Si(111) wafers (boron-doped) with room temperature resistivities of 0.01 $\Omega\cdot$cm and 0.001 $\Omega\cdot$cm, respectively, were used as substrates for the $\sqrt{3}$-Sn growth. Clean Si(111)-7 $\times$ 7 surfaces were obtained by repeatedly flashing well-degassed substrates at 1200$^\textrm{o}$C. The $\sqrt{3}$-Sn phase was then prepared by evaporating  high-purity (99.9999\%) Sn from a standard Knudsen cell on Si(111) at approximately 550$^\textrm{o}$C. All STM measurements were conducted at 4.2 K using sharp polycrystalline PtIr tips, which were routinely cleaned by electron beam bombardment in MBE and appropriately calibrated on MBE-prepared Ag/Si(111) films. The STM topographies were acquired in a constant-current mode with the bias $V$ applied to the sample, while the tunneling spectra were measured using a standard lock-in technique with a small bias modulation at 919 Hz.
\end{spacing}

\begin{figure}[t]
\includegraphics[width=1\columnwidth]{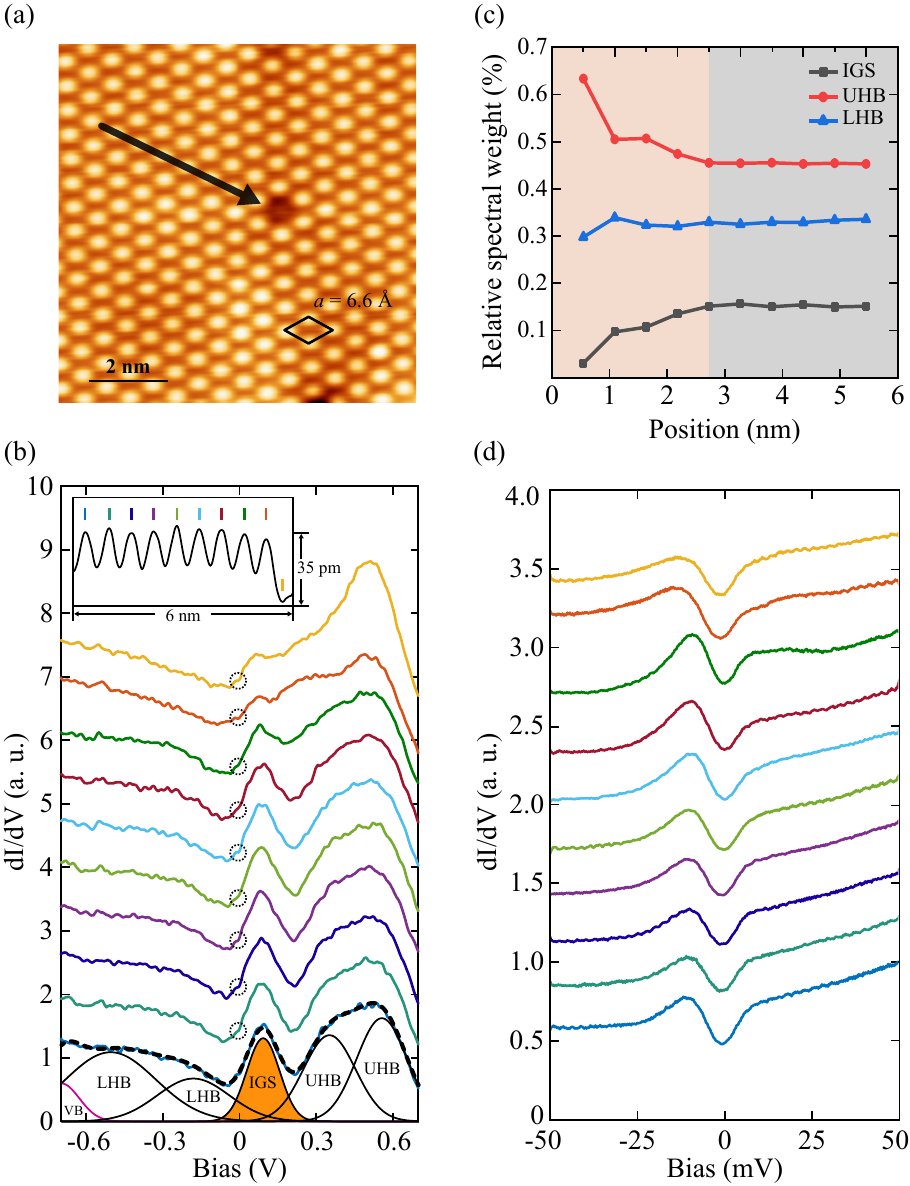}
\caption{(color online) (a) STM topography of $\sqrt{3}$-Sn surface (9 nm $\times$ 9 nm, $V$ = 200 mV, $I$ = 100 pA) with a single Sn vacancy appearing as a dark spot. (b) Large-energy-scale \textit{dI/dV} conductance spectra measured at equal separations (0.66 nm) along the black arrow in (a), color-coded to match the vertical bars (mark the positions for acquiring the \textit{dI/dV} spectra) in the inserted line profile. The bottom spectrum is illustratively fitted with multiple Gaussian peaks as denoted. Tunneling junction is stabilized at $V$ = 1.0 V and $I$ = 150 pA. (c) Relative spectral weights of the LHB, UHB, and IGS, calculated from their site-specific Gaussian distributions. (d) Low-energy-scale tunneling spectra ($V$ = 50 mV, $I$ = 150 pA) at the same positions as (b), showing a PG opening at $E_\textrm{F}$.
}
\end{figure}

\begin{figure}[t]
\includegraphics[width=\columnwidth]{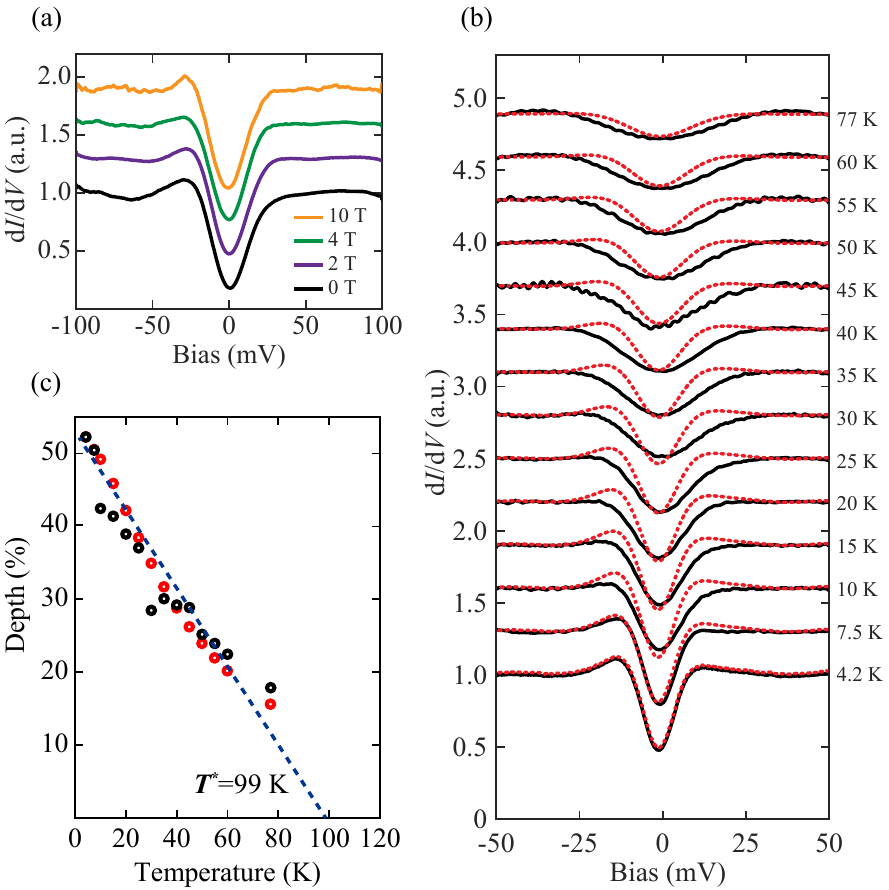}
\caption{(color online) (a) Tunneling conductance \textit{dI/dV} spectra ($V$ = 100 mV, $I$ = 150 pA) of the $\sqrt{3}$-Sn phase under various magnetic fields as indicated. (b) Variation of PG with temperature ($V$ = 50 mV, $I$ = 150 pA). The dotted red lines represent the thermal broadening of the spectrum at 4.2 K to the progressively increasing temperatures. (c) Plotted PG depth as a function of temperature. The black and red circles represent the values extracted from the experimental curves and the thermal broadening ones in (b), respectively, which are nicely matched to each other. The dashed line shows the best fit of PG depth to a linear function.
}
\end{figure}

\begin{figure*}[t]
\includegraphics[width=2\columnwidth]{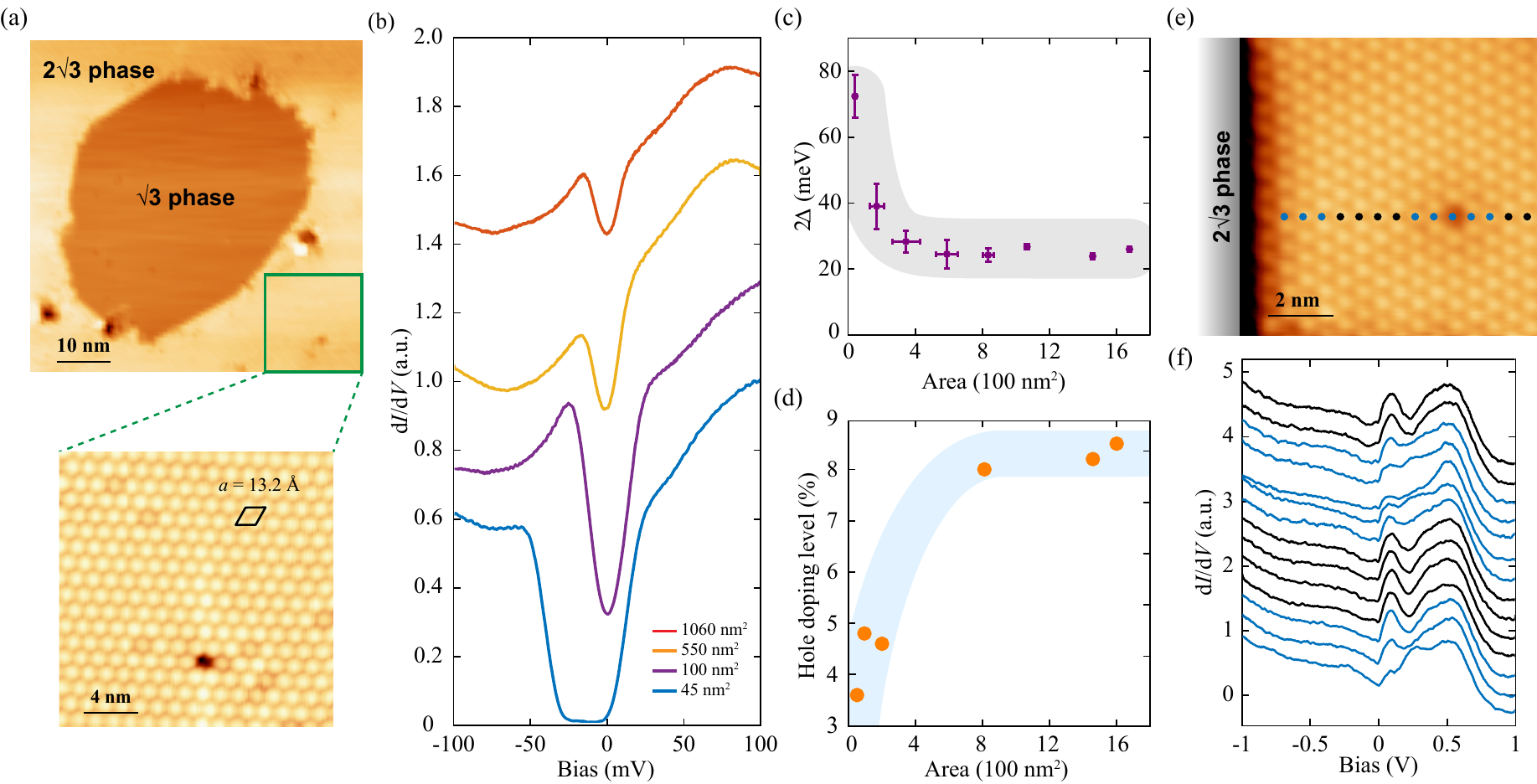}
\caption{(color online) (a) STM topography (60 nm $\times$ 60 nm, $V$ = 2.0 V, $I$ = 90 pA) showing an isolated $\sqrt{3}$-Sn domain encircled by semiconducting $2\sqrt{3}$-Sn phase. The bottom panel shows an atomically resolved STM image of the $2\sqrt{3}$-Sn surface (20 nm $\times$ 20 nm, $V$ = -0.7 V, $I$ = 150 pA). (b) Spatially averaged \textit{dI/dV} spectra on heterogenous $\sqrt{3}$-Sn domains with varying areas as indicated. The spectra have been vertically offset for clarity. Tunneling junction is stabilized at $V$ = 100 mV and $I$ = 150 pA. (c, d) Dependence of the PG magnitude and hole doping level on the area of $\sqrt{3}$-Sn domains, respectively. The error bars in (c) denote the standard deviations of the PG magnitude and area from various $\sqrt{3}$-Sn domains. (e) STM topography (9 nm $\times$ 9 nm, $V$ = 50 mV, $I$ = 100 pA) across the boundary between the $\sqrt{3}$-Sn and $2\sqrt{3}$-Sn phases. (f) Tunneling spectra ($V$ = 1.0 V, $I$ = 150 pA) measured at the dot-marked sites, color-coded to match the dots in (e).
}
\end{figure*}

Figure 1(a) shows a representative STM topography of $\sqrt{3}$-Sn prepared on the 0.001 $\Omega\cdot$cm Si substrates, with the bright spots denoting the Sn adatoms. They occupy the $T_\textrm{4}$ sites of the underlying Si (111) surface and are spaced $\sim$ 6.6 $\textrm{\AA}$ apart. Individual Sn vacancies occur as dark contrast that change little with the sample bias. Figure 1(b) plots a series of differential conductance \textit{dI/dV} spectra (from bottom to top) taken along a trajectory (marked by the black arrow) approaching a single Sn vacancy in Fig.\ 1(a). As anticipated, the Mott Hubbard bands could be readily identified spectroscopically. In line with a previous report \cite{ming2017realization}, every \textit{dI/dV} spectrum can be nicely decomposed into six Gaussian peaks (see the best fit of the bottom one to multiple Gaussian function), derived from the slightly appearing valence band (VB), lower Hubbard band (LHB), quasiparticle peak or saying in-gap states (IGS) at around 100 meV, and upper Hubbard band (UHB), respectively. As justified before, the hole doping level of $\sqrt{3}$-Sn can be estimated by calculating the spectra weight of IGS relative to those of UHB and LHB in Fig.\ 1(c) \cite{ming2017realization}. Apparently, the IGS become less prominent in the vicinity of Sn vacancy, indicative of a reduced hole doping level there. It is understandable because one missing Sn adatom breaks three Sn-Si bonds, leaving behind three Si dangling bonds. This contributes two extra electrons and reduces the local hole doping. Consequently, the spectral weight of UHB increases as a compensation of the significantly suppressed IGS around the Sn vacancy [Fig.\ 1(c)]. Such behavior resembles those of both hole- and electron-doped cuprates, in which the IGS also emerge upon doping \cite{zhong2020direct,wang2020electronic,hu2021momentum}. A major difference is the occurrence of more prominent IGS in $\sqrt{3}$-Sn, which might originate from the stronger interaction between the $\sqrt{3}$-Sn and underlying Si(111) substrates.

Another noteworthy feature is a small dip that universally occurs near $E_\textrm{F}$, circled in Fig.\ 1(b). To present the feature more clearly, the smaller-energy-scale \textit{dI/dV} spectra are acquired at the correspondingly identical sites and plotted in Fig.\ 1(d). A spatially uniform energy gap of approximately 20 meV, irrespective of the existence of Sn vacancy, is identified and symmetric with respect to $E_\textrm{F}$. The results are consistent with the previous findings, except that no apparent van Hove singularity is observable below $E_\textrm{F}$ in our study \cite{ming2018zero,ming2017realization}. The robustness of the small energy gap against the Sn vacancy hints at its delocalized nature.

To provide more insights into the observed energy gap, we characterize its dependence on the external magnetic field and temperature in Fig.\ 2. Firstly, the gap remains essentially unchanged upon the application of a magnetic field up to 10 T [Fig.\ 2(a)]. This hints that it has nothing to do with superconductivity. Indeed, the gap magnitude of $\sim$ 20 meV turns out to be unrealistically larger than the specific superconducting gap of $\sim$ 1.5 meV identified in this system \cite{wu2020superconductivity}. Figure 2(b) plots the dependence of the site-specific tunneling \textit{dI/dV} spectra on temperature ranging from 4.2 K to 77 K. At elevated temperatures, the gap gets gradually suppressed. To elucidate the temperature effect, we calculate the convolution curves of the \textit{dI/dV} spectrum measured at 4.2 K with a Fermi-Dirac function at varying temperatures and overlay them on the corresponding \textit{dI/dV} spectra by the red dashed lines. Apparently, they fit nicely to each other especially for the gap depth [Fig.\ 2(c)], defined as the difference between unity and the normalized zero-bias conductance. This is reminiscent of the PG, which also only smears out at elevated temperatures via thermal broadening, in underdoped cuprates \cite{renner1998pseudogap,fischer2007scanning}. By extrapolating the linear relationship of the PG depth versus temperature to the point where the gap depth is equal to zero, the transition temperature, $T^*$, is estimated to be around 99 K. We further compare the spectral weight of IGS relative to LHB and UHB, and therefore obtain the hole doping level of at most 8.5\% for our samples investigated. This value is lower than the threshold doping level of 10\% required for the emergent superconductivity \cite{wu2020superconductivity}. This suggests that the $\sqrt{3}$-Sn samples studied here lie in the underdoped regions, which happens to match with the experimental observation of PG.

It should be noted that such PG feature has been alternatively explained as a dynamical Coulomb blockade effect \cite{ming2018zero}. Given that several different Si(111) substrates with varying doping levels have been explored and the resistive depletion layer between the $\sqrt{3}$-Sn domains and the substrates has been largely modified, the PG feature should profoundly change with the varying substrates in the Coulomb blockade model. This contradicts with the universal occurrence of PG feature in $\sqrt{3}$-Sn, even when the underlying Si(111) substrates are more lightly doped with a resistivity of 0.01 $\Omega\cdot$cm (not shown). More importantly, if only the Coulomb blockade comes into play, the accompanying Coulomb staircase behavior commonly occurs as discrete peaks beside the Coulomb gap, which has never been observed. Thus, it seems more tempting to attribute the dip-like feature to the PG opening at $E_\textrm{F}$, as predicated from a quantum Monte Carlo simulation of the 2D Hubbard model \cite{moukouri2000many}.

In order to corroborate this opinion, we have explored the doping evolution of PG by taking advantage of the $\sqrt{3}$-Sn nanodomain size-dependent hole doping level, albeit unknown cause \cite{ming2018zero}. Here the $\sqrt{3}$-Sn nanodomains are isolated and laterally encircled by the neighboring Sn/Si(111)-($2\sqrt{3}\times 2\sqrt{3}$)\textit{R}30$^\textrm{o}$ (shorten as $2\sqrt{3}$-Sn) phase, as illustrated in Fig.\ 3(a). Figure 3(b) represents the characteristic \textit{dI/dV} spectra, disclosing well-defined PG on various $\sqrt{3}$-Sn nanodomains with varying areas. The PG shrinks in magnitude with increasing $\sqrt{3}$-Sn nanodomain area and saturates to about 20 meV above 600 nm$^2$ [Fig.\ 3(c)]. In extremely small $\sqrt{3}$-Sn nanodomains, a U-shaped gap develops with an apparent electron-hole asymmetry, analogous to the previous study \cite{ming2018zero}. A sharp and significant distinction is that the hole doping level, estimated from the relative spectral weight analysis just as Fig.\ 1(b), decreases as the area of $\sqrt{3}$-Sn nanodomains is reduced [Fig.\ 3(d)]. This has been unambiguously confirmed as well by exploring the spatial-dependent \textit{dI/dV} spectra perpendicular to the boundary between the $\sqrt{3}$-Sn and $2\sqrt{3}$-Sn phases in Figs.\ 3(e) and 3(f). Similar to those near a single Sn vacancy, the tunneling conductance \textit{dI/dV} spectra near the phase boundary are characteristic of significantly weaken IGS. This means a reduced hole doping level along the phase boundary and thus for the small $\sqrt{3}$-Sn nanodomain as a whole. The large discrepancy from the previous study might arise from either the distinct sample preparation details or the different measurement temperatures. In the previous study, the large- and small-energy scale \textit{dI/dV} spectra are separately measured at 77 K and 4.4 K, which renders a direct comparison of them probably questionable. On the other hand, we have performed all the spectroscopic \textit{dI/dV} measurements at the identical temperature of 4.2 K. This makes the present findings more convincing.

\begin{figure}[t]
\includegraphics[width=0.70\columnwidth]{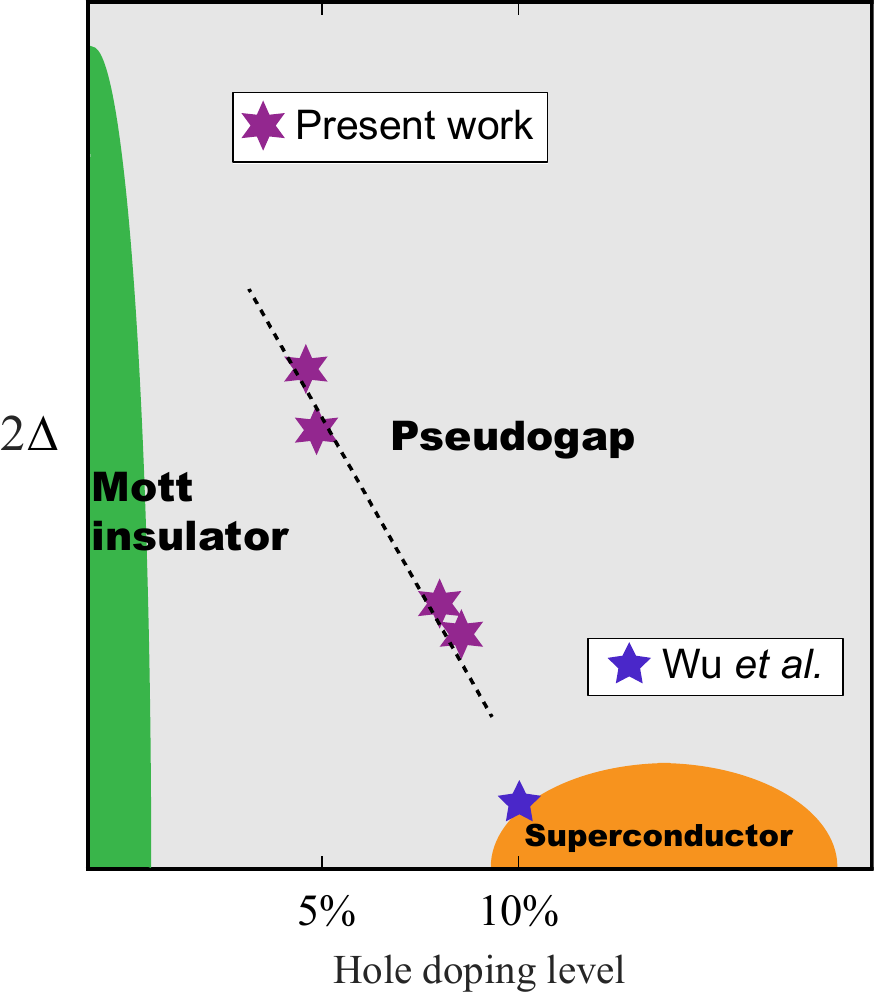}
\caption{(color online) Schematic electronic phase diagram of the modulation-doped $\sqrt{3}$-Sn by the underlying Si(111) substrates. The doping-dependent PG data are acquired on various $\sqrt{3}$-Sn domains in three samples, while the superconductivity one is adapted from Ref.\ 27.
}
\end{figure}

Figure 4 summarizes the doping-dependent magnitude of PG (purple stars) and the superconducting energy gap (blue star) in $\sqrt{3}$-Sn. All PG data are readily extracted from various nanodomains of $\sqrt{3}$-Sn supported by the heavily-doped Si(111) substrates (0.001 $\Omega\cdot$cm). It should be emphasized that we have also measured the $\sqrt{3}$-Sn on a lightly doped Si(111) substrate with the resistivity of 0.01 $\Omega\cdot$cm, which reveals a similar PG except for significantly suppressed gap edge peaks. These behaviors resemble those of cuprate PG in a prominent manner, and the nanodomain size-dependent PG behavior is also expected for a doped 2D Mott insulator \cite{moukouri2000many}.

In summary, we have spectroscopically observed an unusual PG and established the electronic phase diagram in a modulation-doped Mott insulator $\sqrt{3}$-Sn on the Si(111) substrates. The PG, electronic phase diagram, and chemical doping scheme are found to be analogous to those of the high-$T_\textrm{c}$ cuprate superconductors. Further chemical doping control and experimental investigation of $\sqrt{3}$-Sn might find the microscopic mechanism of PG and superconductivity in this doped Mott insulator, by implications in cuprates.

\begin{acknowledgments}
The work is financially supported by the National Natural Science Foundation of China (Grants No.\ 62074092, No.\ 11604366, No.\ 11774192, No.\ 11634007) and the Ministry of Science and Technology of China (Grants No.\ 2018YFA0305603, No.\ 2017YFA0304600).

\end{acknowledgments}

%

\end{document}